\RequirePackage{fix-cm}   
\documentclass[letter,a4paper]{aa}

\usepackage{graphicx}

\long\def\beginpgfgraphicnamed#1#2\endpgfgraphicnamed{\includegraphics{#1}}

\usepackage{microtype}
\usepackage{fixltx2e}

\usepackage[centertags,intlimits]{amsmath}
\usepackage{amssymb,amsfonts}
\usepackage{mathtools}\mathtoolsset{centercolon}

\usepackage{txfonts}

\usepackage{natbib}
\usepackage[breaklinks,pdfborder={0 0 0}]{hyperref}

\begin{document}

\title{Accurate rate coefficients for\\ models of interstellar
  gas-grain chemistry}
\titlerunning{Accurate rate coefficients for models of gas-grain chemistry}
\date{\today}

\author{Ingo Lohmar\inst{1} \and Joachim Krug\inst{1} \and Ofer Biham\inst{2}}
\institute{Institute for Theoretical Physics, 
University of Cologne, Cologne, Germany
\and Racah Institute of Physics, 
The Hebrew University, Jerusalem 91904, Israel}

\begin{abstract} 
{}{The methodology for modeling grain-surface
chemistry has been greatly improved by
taking into account the grain size and fluctuation effects.  
However, the reaction rate coefficients currently used in 
all practical models of gas-grain chemistry are inaccurate
by a significant amount.
We provide expressions for these crucial rate
coefficients that are both accurate and easy
to incorporate into gas-grain models.
}
{We use exact results obtained in earlier work,
where the reaction rate coefficient
was defined by a first-passage problem, which was solved using
random walk theory. 
}
{The approximate reaction rate coefficient presented here
is easy to include in all models of interstellar gas-grain
chemistry.
In contrast to the commonly used expression, 
the results that it provides are
in perfect agreement with detailed kinetic Monte Carlo simulations.
We also show the rate coefficient for reactions involving multiple
species.
}
{}
\end{abstract}


\maketitle

\section{Introduction}

The chemistry in interstellar clouds of gas and dust is very complex,
consisting of reactions taking place in the gas phase as well as
on the surfaces of dust grains 
\citep{herbst95}.  
Of the latter reactions, the formation of molecular hydrogen 
\citep{hollenbach71b}
is of particular importance, but many more processes have been
identified 
\citep{hasegawa92}.  
Modeling of these complex gas-grain chemical networks 
is important to the understanding of the observed
chemical complexity in interstellar clouds and the 
effects of those molecules on gravitational collapse
and star formation
\citep{herbst95,hartquist95,tielens05}.

While for gas-phase reactions,
rate equations are widely used and fully appropriate,
for grain surface chemistry they were found to
be inaccurate for a wide range of astrophysically
relevant conditions 
\citep{tielens95,caselli98,biham02,biham05,lohmar06,lederhendler08}.
To address this problem, modified rate equations were proposed
\citep{caselli98,stantcheva01,garrod08}.
They were found to provide improved results,
but involve ad-hoc and uncontrolled approximations.
The master equation, describing the time evolution in the
probability distribution of the number of reactants on the grain, 
provides a complete and accurate description of the reactions
on grain surfaces
\citep{biham01,green01,biham02}.

The surface chemistry was also studied using
stochastic simulations
\citep{tielens82,charnley01}.
Kinetic Monte Carlo simulations (KMC) have been developed
for different surface morphologies
\citep{chang05,cuppen05}.

An important advantage of approaches based on 
differential equations is that they can easily be 
coupled to the rate equations of gas-phase chemistry.
The difficulty with the master equation is that the number
of equations quickly proliferates as the number of reactive
species on the grains increases, thus approximations are needed
\citep{stantcheva02,stantcheva03}.  
From the master equation,
it is possible to construct a set of moment equations
that account correctly for the
reaction rates on grain surfaces
\citep{lipshtat03,barzel07a,barzel07b}.  

Here we consider the form of the rate coefficients that enter the models.  
Apart from KMC simulations, all approaches to grain-surface chemistry
feature the \emph{sweeping rate} $A$ as a crucial parameter.
This rate coefficient governs the reaction rate
on the surface,
where there is competition between diffusion-mediated
reaction and desorption.
Without a precise
definition of $A$, 
the expression $A=a/S$
is widely used, where $a$ is the hopping rate 
between adjacent adsorption sites 
and $S$ is the number of these sites on the grain.
This expression does not account for the competition 
between reaction and desorption.
It also does not account for 
the peculiarities of two-dimensional diffusion.

\citet{lohmar06} 
introduced a proper definition of the rate coefficient 
$A$ in terms of the two-particle 
\emph{encounter probability}. 
These authors
evaluated it exactly for
a single-species reaction on 
a closed two-dimensional surface 
where the adsorption sites form a regular lattice
\citep{lohmar06,lohmar08}.
The result was compared to the conventional approximation
for $A$, which was found to deviate significantly.
However, the exact expression contains a large sum of terms,
which is costly to evaluate in practice.

In this Letter, we present an expression for the
reaction rate coefficient
that is both accurate and easy to 
evaluate.
We compare the reaction rates obtained using this
expression to those obtained from KMC simulations
and find perfect agreement.
We also generalize the expression to the case
of reactions involving multiple species.

\section{Review of earlier results}
\label{sec:review}

There is no need here to elaborate on the various analytical
approaches to grain-surface reaction kinetics, which we cited before.
Throughout, we consider a system of $S$ adsorption sites arranged in a
two-dimensional quadratic square lattice with periodic boundaries,
onto which atoms impinge at a certain homogeneous flux $F=fS$.  They
perform nearest-neighbor hops with (undirected) rate $a$ and desorb at
a rate $W\ll a$.  The sweeping rate $A$ then appears in all approaches
as the pre-factor in the recombination term and the corresponding
production rate.  For example, the rate equation for the mean number
of particles on the grain, $\langle N\rangle$, reads $\mathrm d\langle
N\rangle/\mathrm d t =F -W\langle N\rangle -2A\langle N\rangle^2$,
with a production rate $A\langle N\rangle^2$ of molecules.  In the
master equation and the moment equations, the latter rate becomes
instead $R =A (\langle N^2\rangle -\langle N\rangle)$
\citep[e.g.]{biham02,lipshtat03}, where the moments $\langle
N^k\rangle =\sum_{N=0}^\infty N^kP(N)$, and $P(N)$ is the probability
of having $N$ particles on the grain at a given time.  Any adsorbed
atom may end up either reacting (with rate $A$) or desorbing (with
rate $W$).  Therefore, the probability for an atom to end up reacting
is given by $p=A/(A+W)$.  With this, we can define $A$ by means of
\citep{lohmar06}
\begin{equation}
  \label{A-def}
  \frac{A}{W} = \frac{p}{1-p},
\end{equation}
where $p$ is the \emph{encounter probability} that two specific
particles present on the grain meet before one of them desorbs.  If
atoms only react with a certain probability upon encounter, $p$ is
simply multiplied by this probability.  
In Sect.~\ref{sec:two-species} below, we derive the corresponding relation
in the context of multi-species reactions, and also show how the
encounter probability generalizes to that case.

The encounter probability has been studied in great detail
\citep{lohmar06,lohmar08}.  Generally, it features two regimes: the
`small-grain' regime, where $SW/a\ll1$ and encounter is almost sure,
$1-p\ll1$, and the `large-grain' regime, where $SW/a\gg1$ and
encounters are rare, $p\ll1$.  The best model at hand is given by a
discrete-time random walk analysis.  Well-known results then yield $p$
in terms of a single sum, and the asymptotic behavior is given by
\citep{lohmar08}
\begin{equation}
  p \approx 1-\frac{SW}{a}\frac{\ln(cS)}{\pi}
\end{equation}
(with a numerical constant $c=1.8456047840\dots$) in the small-grain
regime, while in the large-grain regime,
\begin{equation}
  \label{large-grain-p}
  p\approx \frac{a}{SW}\frac{\pi}{\ln(8a/W)}.
\end{equation}
Using Eq.~\eqref{A-def} thus yields the small-grain asymptotic
\begin{equation}
  \label{small-grain-A}
  A\approx\frac{W}{1-p} \approx \frac{a}{S}\frac{\pi}{\ln(cS)},
\end{equation}
and for large grains,
\begin{equation}
  \label{large-grain-A}
  A\approx Wp \approx \frac{a}{S}\frac{\pi}{\ln(8a/W)}.
\end{equation}
This behavior is largely independent of the precise model used to
obtain $p$, the latter only affecting the pre-factor inside the
logarithm \citep{lohmar08}.
The logarithmic factor
in Eqs.~\eqref{small-grain-A}
and~\eqref{large-grain-A}
reflects the return probability of a two-dimensional random
walker (``back diffusion'') and as such, is also typical of other
two-dimensional diffusion problems
\citep{krug03}.

\section{Practical approximation}
\label{sec:practical-A}

The exact expressions for $p$ are typically complicated, but the
asymptotic results for the sweeping rate of interest are fairly
simple.  Moreover, its functional form is the same in both regimes,
only differing in the argument to the logarithm, which naturally
suggests an interpolation formula.

We first considered to simply use the minimal argument to the
logarithm, making the dependence continuous and producing the correct
limiting behavior in both regimes.  While this appears to be natural
and is very easy to implement, it also introduces a cusp to the
dependence, and it is an arbitrary choice: Any function of $cS$ and
$8a/W$ that approaches the smaller argument in the corresponding limit
could be used as well.  We thus tested $ \left[(cS)^{-n}
  +(8a/W)^{-n}\right]^{-1/n}$ instead.  Evidently, this has the proper
limits for $n>0$, and the exponent can be used to tune the behavior
in-between; the minimum function corresponds to $n\to\infty$.  For
several values of $a/W\gg1$, we ``visually'' tuned the exponent to
yield the smallest relative deviation from the exact values, and ended
up at values $n=1.07\dots1.08$, which decrease the maximum deviation
with respect to the exact result from between $3$--$7\%$ (for
the min function) to $\sim1/30$th of this value.  Numerically, this
however is an extraordinary complication.  Being so close to unity,
$n=1$ is a feasible choice instead, and this still decreases the
maximum deviation by a factor of about $1/6$ in the examples shown
below.  Consequently, we propose the approximation
\begin{equation}
  \label{A-approx}
  A \approx \frac{a}{S}\cdot\pi
  \left[ \ln \frac{1}{1/(cS)+W/(8a)} \right]^{-1},
\end{equation}
where $c=1.8456047840\dots$.  
This is the main result of this Letter,
and we advocate its use in all models of grain-surface chemistry.

\subsection{Comparison}

\begin{figure}[htbp!]
  \centering
  \beginpgfgraphicnamed{12746fg1}
  \begin{tikzpicture}
    \begin{semilogxaxis}[smooth,xlabel=$S$,ylabel=$\frac{A}{a/S}$,
      width=\columnwidth,height=7cm,
      xmin=60,xmax=1.9e9,ymin=0,ymax=0.64,
      y tick label style={/pgf/number format/precision=1},
      axis x line=bottom,axis y line=left,xlabel style={above},
      skip coords between index={36}{99}]
      \pgfplotstableread{tbcrw.dat}\table;
      \addplot[thick] plot table[x index=0,y index=1] from \table
      node[above left] {$W/a=10^{-3}$};
      \addplot[thick] plot table[x index=0,y index=2] from \table
      node[above left] {$10^{-4}$};
      \addplot[thick] plot table[x index=0,y index=3] from \table
      node[above left] {$10^{-6}$};
      \addplot[only marks,mark=o] plot table[x index=0,y index=4] from \table;
      \addplot[only marks,mark=o] plot table[x index=0,y index=5] from \table;
      \addplot[only marks,mark=o] plot table[x index=0,y index=6] from \table;
    \end{semilogxaxis}
  \end{tikzpicture}
  \endpgfgraphicnamed
  \caption{Ratio between the correct sweeping rate $A$ and the
    conventional approximation $a/S$, as obtained from the exact
    expression for $A$ (solid lines) and from the approximate form
    (circles) given by Eq.~\eqref{A-approx}, for three different
    values of $W/a\ll1$.}
\label{fig:A-reduction}
\end{figure}
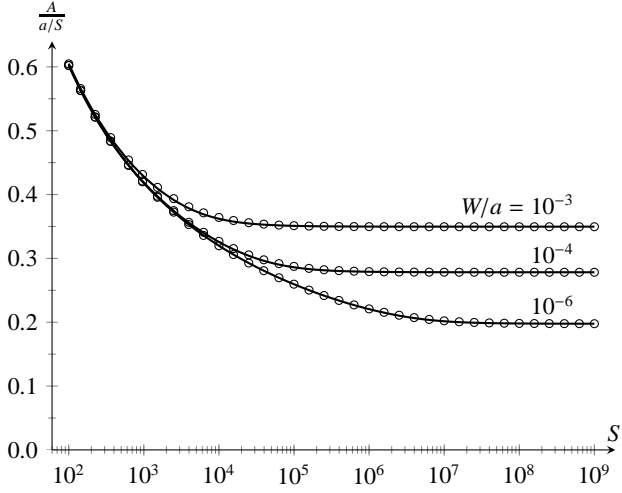

In Fig.~\ref{fig:A-reduction}, we present the exact sweeping rate
together with the approximate expression in Eq.~\eqref{A-approx}, both
normalized by the conventional approximation $a/S$.
The approximation proposed above is in perfect
agreement with the exact result.  Using thermal activation of rates
with activation energies for amorphous carbon as given by
\citet{katz99}, $W/a=10^{-3}$ occurs at $T\approx21.34\,\mathrm K$,
and the other values at $16.00\,\mathrm K$ and $10.67\,\mathrm K$,
respectively.
The discrepancy with respect to the conventional approximation is
found to be most pronounced for large grains.  We note that even for
the smallest grains, where the correction is least severe, the factor
is still well below unity.

\subsection{Lattice-type effects}
In the above, we considered the case of a quadratic square lattice.
The restriction to the quadratic case, in which the lattice extends to
the same lengths in both dimensions, is not easily removed; indeed,
rectangular lattices exhibit a more complex behavior than desired for
a simple model \citep{lohmar08}.  Different lattice types \emph{can}
be examined rather easily however, and we also considered the
triangular lattice.  For the exact expressions, the effect on the encounter
probability $p$ as well as the sweeping rate $A$ has been shown to be
minute throughout \citep{lohmar06,lohmar08}.  In the approximation
presented herein, the triangular lattice simply amounts to a change in
the pre-factors in Eq.~\eqref{A-approx}, namely
\begin{equation}
  A \approx \frac{a}{S} \cdot \frac{2\pi}{\sqrt{3}}
  \left[ \ln \frac{1}{1/(\tilde cS)+W/(12a)} \right]^{-1},
\end{equation}
where now $\tilde c=2.3472914383\dots$.

\subsection{Langmuir-Hinshelwood rejection}
Expressions for the encounter probability given so far include
coinciding initial positions of both particles.  The question of
whether Langmuir-Hinshelwood rejection should be accounted for on the
`global' level of the surface-chemistry model is beyond the scope of
this Letter; most likely, it is relevant only for H and D atoms.
Irrespective of this, we are interested here in the most accurate
numerical
value for the reaction rate coefficient as opposed to a consistent
theoretical framework (e.g., one may even use rate equations in the
end).  Consequently, one might want to use
\begin{equation}
\tilde p = \frac{Sp-1}{S-1} < p
\end{equation}
for the encounter probability, which does not permit coinciding
initial positions \citep{lohmar08}.  The implied change in the
sweeping rate is given by
\begin{equation}
  \tilde A = \frac{W\tilde p}{1-\tilde p}
  = A - \frac{W}{S(1-p)}.
\end{equation}
We evaluate this for both regimes separately.  For small grains,
$W/(1-p)\approx A$, and thence, $\tilde A\approx A(1-1/S)$.  For large
grains, $W\approx A/p$, such that
\begin{equation}
  \tilde A \approx A \left(1-\frac{1}{Sp(1-p)}\right).
\end{equation}
Here, $p\ll1$, so $p(1-p)\approx p$, and from
Eq.~\eqref{large-grain-p},
\begin{equation}
  \tilde A\approx A\left(1-\frac{W}{a}\frac{\ln 8a/W}{\pi}\right).
\end{equation}
Since both $S$ as well as $a/W$ can safely be assumed to be larger
than a few hundred, the relative corrections to $A$ are below $1\%$,
and this is as good as can be achieved with the interpolation in $A$.
Consequently, accounting for Langmuir-Hinshelwood rejection in the
sweeping rate in practice is unnecessary.

\section{Kinetic Monte Carlo simulations}
\label{sec:kmc}
We now show that using the approximation in Eq.~\eqref{A-approx} in
analytical approaches is in excellent agreement with KMC simulations.
In our context, we restrict ourselves to a proof-of-principle
comparison, hence we treat only the case of single-species
recombination.  We also use the master equation approach as the most
precise model, to avoid discrepancies that might arise from factors
unrelated to our approximations.  In practice, one would use an
approximation such as the moment equations \citep{lipshtat03}, which
have been shown to produce accurate results for a broad range of
conditions, and to be applicable to large-scale astrochemistry
simulations \citep{barzel07a,barzel07b}.

The single-species master equation admits a stationary solution
\citep{green01,biham02}.  
The \emph{efficiency} $\eta$ of the process
is then given by the production rate $R$ (molecules produced on the
grain per unit time) normalized by the influx, and one obtains
\begin{equation}
  \label{me-eta}
  \eta = \frac{R}{F/2}
  = \frac{I_{W/A+1}(2\sqrt{2F/A})}{I_{W/A-1}(2\sqrt{2F/A})},
\end{equation}
where $I_\mu(z)$ is the modified Bessel function of the first kind.

Our KMC simulations use a standard algorithm \citep[described, e.g.,
in][]{chang05}, which has been suitably optimized for the
peculiarities of our system, and the code has been tested extensively.
Atoms impinging onto occupied sites are rejected by the
Langmuir-Hinshelwood mechanism, while we do not account for this
process in the analytical treatment --- this illustrates that the
framework presented here is fully appropriate for moderate coverages
(i.e., for the high-temperature drop in efficiency).  We also wait for
a long enough time to establish steady-state conditions, and follow
the system for $10^8$ impingements.

\begin{figure}[htbp!]
  \centering
  \beginpgfgraphicnamed{12746fg2}
  \begin{tikzpicture}
    \begin{semilogxaxis}[smooth,xlabel=$S$,ylabel=$\eta$,
      width=\columnwidth,height=65mm,
      xmin=75,xmax=1.9e7,ymin=0,ymax=0.245,
      y tick label style={/pgf/number format/precision=2},
      axis x line=bottom,axis y line=left,xlabel style={above}]
      \pgfplotstableread{etas-carbon-18K.dat}\table;
      \addplot[thick] plot table[x index=0,y index=1] from \table
      node[below left] {$A$ as of~(6)}; 
      \addplot[thick,densely dashed] plot table[x index=0,y index=2] from \table
      node[below left] {$A\approx a/S$};
      \addplot[only marks,mark=o] plot file{eta-KMC.dat};
    \end{semilogxaxis}
  \end{tikzpicture}
  \endpgfgraphicnamed
  \caption{Recombination efficiency versus grain size, as obtained
    from KMC simulations (circles), from the master equation using
    Eq.~\eqref{A-approx} (solid line) and using the conventional
    approximation $a/S$ (dashed line).}
\label{fig:eta-comparison}
\end{figure}
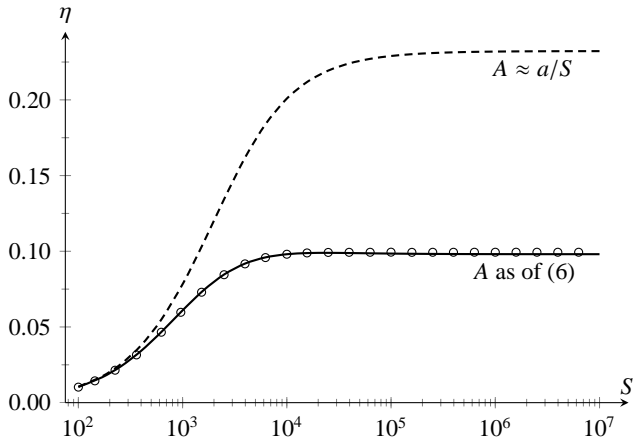

Surface parameters are chosen as found experimentally for amorphous
carbon surfaces \citep{katz99}, with thermal activation energies
$E_W/k_\mathrm{B}\approx698\,\mathrm K$,
$E_a/k_\mathrm{B}\approx511\,\mathrm K$, and an attempt frequency
$\nu=10^{12}\,\mathrm s^{-1}$.  With experimentally determined site
densities and assuming standard gas phase conditions (hydrogen atom
density of $n_\mathrm{H}=10\,\mathrm{cm}^{-3}$ at a temperature of
$100\,\mathrm{K}$), one obtains an H influx
$f=7.3\times10^{-9}\,\mathrm{monolayers}/\mathrm s$ \citep{biham01}.
For these conditions and a grain temperature of $T=18\,\mathrm K$, the
results of the KMC simulations are compared to the master equation
steady-state efficiency in Eq.~\eqref{me-eta} using either the
approximation to $A$ given by Eq.~\eqref{A-approx} or the conventional
$A\approx a/S$ in Fig.~\ref{fig:eta-comparison}.  Evidently, the
agreement is perfect using the approximation of $A$ given in
Eq.~\eqref{A-approx}, while the conventional approximation is in
significant disagreement.

\section{Generalization to multi-species reactions}
\label{sec:two-species}

Our method has been formulated for single-species reactions
exclusively.
As speculated by \citet{lohmar06}, we now show that it can easily
be generalized to multi-species reactions.
We use X and Y to denote two different species of particles.
First, we consider the two-particle problem to find the encounter
probability $p_\mathrm{XY}$ of an X and a Y particle on a
\emph{translationally invariant} lattice.
So far, we have been concerned with two
particles, each with a (undirected) hopping rate $a$ and desorption rate $W$;
this situation can be mapped onto a single walker moving and
desorbing at rates $2a$ and $2W$, respectively. 
Both rates can then be
rescaled to their old values, since only their ratio is relevant
\citep{lohmar06,lohmar08}.  
For two species X and Y, and the corresponding rates 
$a_\mathrm X$, 
$a_\mathrm Y$ 
and 
$W_\mathrm X$,
$W_\mathrm Y$, 
we map to one walker moving with rate 
$a_\mathrm X + a_\mathrm Y$ 
and desorbing at a rate 
$W_\mathrm X+W_\mathrm Y$.  
The rescaling argument is not applicable for arbitrary rates.  
However, it is evident that we may simply substitute each $a$ in the
single-species $p$ (Sect.~\ref{sec:review}) by $a_\mathrm X +a_\mathrm
Y$ now, and each $W$ by $W_\mathrm X+W_\mathrm Y$.  This yields the
encounter probability $p_\mathrm{XY}$ of a given pair of X and Y
particles.

Second, we derive 
the relation between the encounter
probability $p_\mathrm{XY}$ 
and the sweeping rate $A_\mathrm{XY}$, 
following \citet{lohmar06}.  
We start from the production rate of XY molecules,
\begin{equation}
R_\mathrm{XY} = A_\mathrm{XY} \langle N_\mathrm X N_\mathrm Y \rangle,
\end{equation}
in an analogous way to the single-species situation.  
This means that the reaction
reduces the number of XY pairs at a rate $A_\mathrm{XY}$
(compared to a rate $2A$ in the single-species case).
An argument fully analogous to the single-species reaction case infers
the relation to the encounter probability $p_\mathrm{XY}$ of a given
XY-pair.  The encounter probability is given by the ratio of the rate
at which the reaction takes away a pair to the overall rate at which a
(particular) pair is removed, by either the reaction or the desorption
of either constituent,
\begin{equation}
  p_\mathrm{XY} = \frac{A_\mathrm{XY}}{A_\mathrm{XY}
    +W_\mathrm X +W_\mathrm Y}.
\end{equation}
Consequently,
\begin{equation}
  A_\mathrm{XY} = \frac{\left(W_\mathrm X +W_\mathrm Y\right)
    p_\mathrm{XY}}{1-p_\mathrm{XY}}
\end{equation}
is the definition for the reaction rate coefficient of $\mathrm X
+\mathrm Y \longrightarrow\mathrm{XY}$.

Since the right-hand side has the same form as for the single-species
reaction with the substitution $W\to W_\mathrm X+W_\mathrm Y$ (and
hopping rates do not occur here), the sum of desorption rates cancels
with the factor of the $p_\mathrm{XY}$ asymptotics just as it does for
the single-species case.  Using Eq.~\eqref{A-approx}, we thus obtain
the approximation
\begin{equation}
  A_\mathrm{XY} \approx \frac{a_\mathrm X+a_\mathrm Y}{S}
  \pi \left[ \ln\frac{1}{\frac1{cS}
      + \frac{W_\mathrm X+W_\mathrm Y}{8(a_\mathrm X+a_\mathrm Y)}}
    \right]^{-1}
\end{equation}
for the reaction rate coefficient of 
$\mathrm X+\mathrm Y \longrightarrow\mathrm{XY}$,
where $c=1.8456047840\dots$.

\section{Conclusions}

We have used exact results for the sweeping rate in diffusion-mediated
reactions on dust grain surfaces to provide an approximate formula for
the crucial reaction rate coefficient.  This expression is numerically
accurate, and easy and efficient to implement even in complex
gas-grain reaction networks.  For single-species reactions, the
expression was employed in the exact analytical master equation
framework, and comparison with KMC simulations shows excellent
agreement throughout.  The results were also generalized to reactions
of multiple species.  We strongly advocate the incorporation of the
reaction rate coefficients presented here into \emph{all} models of
interstellar gas-grain chemistry.  This could be easily achieved and,
in stark contrast to the commonly used approximation, it provides
accurate results at virtually no additional computational cost.  This
is an important step towards an accurate modeling of the fascinating
chemistry of interstellar clouds.

\begin{acknowledgements}
  This work was supported by DFG within SFB/TR-12 \textit{Symmetries
    and Universality in Mesoscopic Systems}.  J.K.\ acknowledges the
  gracious hospitality of Hebrew University and the Lady Davis
  Fellowship Trust.
\end{acknowledgements}

\bibliographystyle{aa}
\bibliography{jrnlabrv,myrefs}

\begin{thebibliography}{27}
\expandafter\ifx\csname natexlab\endcsname\relax\def\natexlab#1{#1}\fi

\bibitem[{Barzel \& Biham(2007{\natexlab{a}})}]{barzel07a}
Barzel, B. \& Biham, O. 2007{\natexlab{a}}, ApJ Lett., 658, l37

\bibitem[{Barzel \& Biham(2007{\natexlab{b}})}]{barzel07b}
Barzel, B. \& Biham, O. 2007{\natexlab{b}}, J. Chem. Phys., 127, 144703

\bibitem[{Biham {et~al.}(2001)Biham, Furman, Pirronello, \& Vidali}]{biham01}
Biham, O., Furman, I., Pirronello, V., \& Vidali, G. 2001, ApJ, 553, 595

\bibitem[{Biham {et~al.}(2005)Biham, Krug, Lipshtat, \& Michely}]{biham05}
Biham, O., Krug, J., Lipshtat, A., \& Michely, T. 2005, Small, 1, 502

\bibitem[{Biham \& Lipshtat(2002)}]{biham02}
Biham, O. \& Lipshtat, A. 2002, Phys. Rev. E, 66, 056103

\bibitem[{Caselli {et~al.}(1998)Caselli, Hasegawa, \& Herbst}]{caselli98}
Caselli, P., Hasegawa, T.~I., \& Herbst, E. 1998, ApJ, 495, 309

\bibitem[{Chang {et~al.}(2005)Chang, Cuppen, \& Herbst}]{chang05}
Chang, Q., Cuppen, H.~M., \& Herbst, E. 2005, A\&A, 434, 599

\bibitem[{Charnley(2001)}]{charnley01}
Charnley, S.~B. 2001, ApJ, 562, L99

\bibitem[{Cuppen \& Herbst(2005)}]{cuppen05}
Cuppen, H.~M. \& Herbst, E. 2005, Mon. Not. R. Astron. Soc., 361, 565

\bibitem[{Garrod(2008)}]{garrod08}
Garrod, R.~T. 2008, A\&A, 491, 239

\bibitem[{Green {et~al.}(2001)Green, Toniazzo, Pilling, Ruffle, Bell, \&
  Hartquist}]{green01}
Green, N. J.~B., Toniazzo, T., Pilling, M.~J., {et~al.} 2001, A\&A, 375, 1111

\bibitem[{Hartquist \& Williams(1995)}]{hartquist95}
Hartquist, T.~W. \& Williams, D.~A. 1995, The Chemically Controlled Cosmos
  (Cambridge University Press)

\bibitem[{Hasegawa {et~al.}(1992)Hasegawa, Herbst, \& Leung}]{hasegawa92}
Hasegawa, T.~I., Herbst, E., \& Leung, C.~M. 1992, ApJ Supp., 82, 167

\bibitem[{Herbst(1995)}]{herbst95}
Herbst, E. 1995, Annu. Rev. Phys. Chem., 46, 27

\bibitem[{Hollenbach {et~al.}(1971)Hollenbach, Werner, \&
  Salpeter}]{hollenbach71b}
Hollenbach, D.~J., Werner, M.~W., \& Salpeter, E.~E. 1971, ApJ, 163, 165

\bibitem[{Katz {et~al.}(1999)Katz, Furman, Biham, Pirronello, \&
  Vidali}]{katz99}
Katz, N., Furman, I., Biham, O., Pirronello, V., \& Vidali, G. 1999, ApJ, 522,
  305

\bibitem[{Krug(2003)}]{krug03}
Krug, J. 2003, Phys. Rev. E, 67, 065102(R)

\bibitem[{Lederhendler \& Biham(2008)}]{lederhendler08}
Lederhendler, A. \& Biham, O. 2008, Phys. Rev. E, 78, 041105

\bibitem[{Lipshtat \& Biham(2003)}]{lipshtat03}
Lipshtat, A. \& Biham, O. 2003, A\&A, 400, 585

\bibitem[{Lohmar \& Krug(2006)}]{lohmar06}
Lohmar, I. \& Krug, J. 2006, Mon. Not. R. Astron. Soc., 370, 1025

\bibitem[{Lohmar \& Krug(2009)}]{lohmar08}
Lohmar, I. \& Krug, J. 2009, J. Stat. Phys., 134, 307

\bibitem[{Stantcheva {et~al.}(2001)Stantcheva, Caselli, \&
  Herbst}]{stantcheva01}
Stantcheva, T., Caselli, P., \& Herbst, E. 2001, A\&A, 375, 673

\bibitem[{Stantcheva \& Herbst(2003)}]{stantcheva03}
Stantcheva, T. \& Herbst, E. 2003, Mon. Not. R. Astron. Soc., 340, 983

\bibitem[{Stantcheva {et~al.}(2002)Stantcheva, Shematovich, \&
  Herbst}]{stantcheva02}
Stantcheva, T., Shematovich, V.~I., \& Herbst, E. 2002, A\&A, 391, 1069

\bibitem[{Tielens(1995)}]{tielens95}
Tielens, A. G. G.~M. 1995, unpublished

\bibitem[{Tielens(2005)}]{tielens05}
Tielens, A. G. G.~M. 2005, The Physics and Chemistry of the Interstellar Medium
  (Cambridge University Press)

\bibitem[{Tielens \& Hagen(1982)}]{tielens82}
Tielens, A. G. G.~M. \& Hagen, W. 1982, A\&A, 114, 245

\end{thebibliography}

\end{document}